\documentclass[12pt]{elsarticle}

\usepackage{amssymb}
\usepackage{amsmath}
\usepackage{graphicx}
\usepackage{dcolumn}
\usepackage{bm}
\usepackage[utf8]{inputenc}
\usepackage[T1]{fontenc}
\usepackage{booktabs}
\usepackage{mathptmx}
\usepackage{etoolbox}
\usepackage{xcolor}
\usepackage{chemformula}
\usepackage{physics}
\usepackage{longtable}
\usepackage{array}
\usepackage{multirow}
\usepackage{threeparttable}
\newcolumntype{P}[1]{>{\centering\arraybackslash}p{#1}}
\newcommand{\mo}[1]{m_{O_{#1}}}
\newcommand{\mc}{m_C}

\begin{document}

\begin{frontmatter}

\title{Correlation between normal character and Raman intensities of isotopologues of carbon dioxide}

\author[1]{E. Suárez}
\author[2]{C. Ostertag-Henning}
\author[3,4]{M. Carvajal}
\author[1]{R. Lemus\corref{cor1}}
\cortext[cor1]{Corresponding author}
\ead{renato@nucleares.unam.mx}

\address[1]{Instituto de Ciencias Nucleares, Universidad Nacional Autónoma de México, Ciudad de México, 04510, Mexico}
\address[2]{Federal Institute for Geosciences and Natural Resources, Hannover, 30655, Germany}
\address[3]{Departamento de Ciencias Integradas y Centro de Estudios Avanzados en Física, Matemáticas y Computación, Unidad Asociada GIFMAN, CSIC-UHU, Universidad de Huelva, Huelva, 21071, Spain}
\address[4]{Instituto Universitario Carlos I de Física Teórica y Computacional, Universidad de Granada, Spain}

\begin{abstract}
Previous studies on the vibrational structure of carbon dioxide isotopologues have revealed a correlation between the strength of Fermi interactions and a recently defined normal-mode parameter. \textcolor{black}{In this contribution, we investigate how the Raman intensities of the most intense bands evolve across the complete series of carbon dioxide isotopologues at room temperature. To this end, we introduce a general scheme for simulating Raman intensities in linear triatomic molecules, based on the computation of the mean polarizability surface up to cubic terms for all twelve stable species. For the first time, we establish correlations between the degree of normality and the Raman transition intensities, mean polarizability derivatives, and variational Raman shift errors across the CO$_2$ isotopologue series. The observed trends are further analyzed by considering rotational contributions, which may partially account for the deviations observed in the correlations.} Our study is supported by experimental Raman spectra of isotopically enriched samples and theoretical simulations. Although this study focuses on carbon dioxide, similar behavior is expected for other series of isotopologues.
\end{abstract}

\end{frontmatter}

\section{Introduction}
The concept of normal modes is of major importance in vibrational spectroscopy \cite{herzberg,praga,califano}. Normal modes provide the frequency associated with collective motion of the atoms specified by the normal coordinates. Their frequencies provide the fundamental energies of the molecule, establishing  the main resonances at low and medium range energies  besides containing information about the geometry and the electronic structure through the force constants. The  direct approach to obtain the normal modes relies on establishing the Lagrangian in terms of Cartesian coordinates. The use of Cartesian coordinates facilitates their calculation from the  technical point of view, giving the possibility to be obtained by standard programs like GAUSSIAN and MOLPRO, for instance \cite{g16,molpro}. However, using Cartesian coordinates,  a direct physical interpretation of the force constants is not possible, avoiding the access to chemical information. A crucial step to solve this problem consists in introducing internal coordinates  \cite{wilson}, allowing the force constants to be linked to the concept of chemical bond in valence bond theory. Their importance is reflected in the relevant work dedicated to establish an appropriate set of coordinates~\cite{hoy}. In this framework the force constants estimated for one molecule  may be  carried over into another similar molecule. In addition, internal coordinates play a crucial role in establishing the concept of local modes, which was coined from overtone spectroscopy studies of molecules with X--H bonds. In general, molecules with terminal bonds involving large mass differences present a local mode behavior at high energies, experimentally proven by the existence of doublets of energy \cite{quack,lawton,child84,halonen98,jensenLoc}. This kind of systems were extensively studied establishing the connection between the interactions when expressed  in the local and the normal schemes \cite{mills,mills2,valle,gambi,duncan1,duncan2,lehmann2,kellman2}.

The force constants depend on the choice of the internal coordinates. As a result, studies aimed at establishing an optimal set based on topology and symmetry considerations have been proposed \cite{kemal}. Another path of investigation, based on the search for invariants under coordinate transformations, was the introduction of the compliance matrix $\Gamma$ \cite{decius}. However, it faces the problem
of lack of a physical basis and, consequently, of a link to vibrational spectroscopy, a fact that leads to the search \textcolor{black}{of better} alternatives to establish local properties. In this venue, Konkoli and Cremer introduced the adiabatic internal coordinate modes \cite{konkoli1,konkoli2,konkoli3,konkoli4}, in which a molecular fragment $\phi_n$ associated with a local coordinate $q_n$ is frozen at its equilibrium while the others are relaxed. This perspective  permits to define  molecular properties  such as  local mode force constant, local mass, local frequency and local mode infrared intensity. A relevant result is that the obtained local force constants reflect the intrinsic strength of the bond, making them suited to be a descriptor of both covalent and noncovalent bonds. In addition, given the relation between local and normal modes, they derived the contribution  of a local mode  to a given normal mode. A remarkable  advantage of this theory is that all this information is obtained using  the standard normal mode analysis, albeit providing a new way to interpret vibrational spectra \cite{kraka1,kraka2}.

More recently, a new criterion to  establish a normal mode degree of a molecule has been proposed based on internal coordinates \cite{marisolpoly2,suarez1}. The criterion relies on  the connection between the normal and local coordinates, albeit using a second quantization scheme. In this way the creation bosonic operator belonging to  a given normal mode is expanded in terms of local creation and annihilation bosonic operators. An average \textcolor{black}{of} the coefficients associated with the annihilation operators provides a parameter $\delta_-$ measuring the degree of normality, which is quite different from the traditional parameter \textcolor{black}{introduced by Child and Halonen \cite{lawton,child84,halonen98,child2}. The connection has been established for pyramidal molecules characterized by local-mode behavior \cite{oscar2023}. It should be remarked that in the traditional local-mode approach the connection between normal and local operators assumes $\delta_-=0$ \cite{halonen98,valle}.}

In order to establish the validity of the normal mode parameter, a detailed analysis was carried out involving both stretching and bending degrees of freedom, even when redundant coordinates are present \cite{suarez1}.  The relevance of this formalism is that both the parameter $\delta_-$ and the coefficients $c^{(\iota,j)}_\pm$ associated with the $j$-th local oscillator have been found to correlate with the interaction strengths and principal components of the vibrational eigenstates \textcolor{black}{for the} isotopologues of carbon dioxide \cite{marisol24}, cyanogen fluoride \cite{suarezfcn}, and carbonyl sulfide \cite{suarezocs}. These correlations were obtained from vibrational descriptions of spectroscopic quality using an algebraic approach based on the $su_1(2) \oplus su(3) \oplus su_2(2)$ dynamical algebra. Importantly, the coefficients governing the most intense Raman transitions also appear closely related to $\delta_-$, suggesting that this parameter may connect the underlying vibrational structure with experimentally measurable observables. 

Although a normal mode analysis involves only second-order force constants, it encodes information reflected in molecular observables, such as infrared and Raman spectra, which depend on multiple factors. In this context, the adiabatic internal modes formalism has been applied to the study of IR intensities \cite{Zou}, and in the present work we extend this approach to Raman spectroscopy, examining the correlation of the normal mode character, as quantified by $\delta_-$, with transition intensities, mean polarizability derivatives, and Raman shifts from variational calculations for a series of carbon dioxide isotopologues. This analysis aims to establish a clear link between \textcolor{black}{the normal} mode behavior and Raman spectroscopic signatures.

\textcolor{black}{In this work, the various isotopologues of carbon dioxide are denoted using the abbreviated AFGL convention \cite{mcclatchey1973,rothman1981}, as employed in the AFGL and HITRAN databases. Within this notation, \ch{^{12}C} and \ch{^{13}C} are represented by 2 and 3, respectively, while \ch{^{16}O}, \ch{^{17}O}, and \ch{^{18}O} are labeled as 6, 7, and 8. Accordingly, a carbon dioxide molecule composed of one \ch{^{13}C} atom and two oxygen atoms, \ch{^{17}O} and \ch{^{18}O}, is written as 738. In total, carbon dioxide comprises 12 stable isotopologues.}

This paper is organized as follows. In Section~\ref{sec-coord},  our selection of the nuclear coordinates is discussed.
Section~\ref{sec-localnormal} is devoted to provide the basic argument leading to establish the normal character of a molecule. \textcolor{black}{The salient features of the vibrational description are presented in Section~\ref{sec-vibdesc}.} In Section~\ref{sec-raman}, a summary of the model to simulate the Raman spectra is given, while in Section~\ref{sec-exp} a description of the experiment is presented. The correlation results supported by both theoretical and experimental Raman intensities are discussed in Section~\ref{sec-results}. Finally, in Section~\ref{sec-conclusions}, the summary and conclusions are drawn.

\section{Nuclear displacement coordinates}
\label{sec-coord}

For linear triatomic molecules we use as \textcolor{black}{curvilinear} internal coordinates the relative distances $q_i=\Delta r_i=r_i-r_{ie}$ for the stretches and the set
\begin{equation}
	\label{qb}
	\textcolor{black}{q}_{2a}=\sqrt{r_{1e} r_{2e}} \ {\bf e}_x\cdot \frac{{\bf r_1}\times {\bf r_2}}{r_1r_2}, \ \ \   \ \ \  \textcolor{black}{q}_{2b}=-\sqrt{r_{1e} r_{2e}} \ {\bf e}_y\cdot \frac{{\bf r_1}\times {\bf r_2}}{r_1r_2} .
\end{equation}
for bends \cite{hoy}. In the particular case of CO\textsubscript{2} isotopologues, it is convenient to introduce the \textcolor{black}{symmetrized curvilinear} coordinates \cite{marisol24}
\begin{equation}
	\label{qt}
	S_{g/u}=\frac{1}{\sqrt{2}}(q_1 \pm q_2); \ \ \ \  \ \ \ \ \  { S}_{\pm}=\mp\frac{1}{\sqrt{2}}\left(\textcolor{black}{q}_{2a} \pm i \textcolor{black}{q}_{2b}\right) ,
\end{equation}
where  ${ S}_\pm$ carry the angular momentum projection. Coordinates $S_{g/u}$, which span irreducible representations $\Sigma_g^+/\Sigma_u^+$ of $\mathcal{D}_{\infty h}$, have been chosen because  the linear terms of their expansion in Cartesian coordinates coincide with the normal coordinates for the symmetric isotopologues. We now define a set of isotopic dependent curvilinear coordinates through the diagonalization of the ${\bf GF}$ matrix, a procedure that imposes the condition
\begin{equation}
	\label{GF}
	{\bf L}^{-1} {\bf G}_0({\bf S}) {\bf F}({\bf S}) {\bf L}={\pmb\Lambda},
\end{equation}
\textcolor{black}{where the Wilson ${\bf G}_0({\bf S})$ and force constant ${\bf F}({\bf S})$ matrices are expressed in terms of the symmetry-adapted coordinates $\tilde{\bf S}=(S_g, S_u)$, and the elements of the diagonal matrix are directly related to the normal-mode frequencies through $\Lambda_{\iota\iota}=\Omega_\iota^2$}. The subscript index in ${\bf G}_{0}$ means that it is calculated at equilibrium. The symmetry adapted   coordinates are defined in terms of the internal coordinates by the transformation $\tilde {\bf S}= \tilde {\bf q} {\bf T}$, and consequently
\begin{equation}
	{\bf G}_0({\bf S})=\tilde {\bf T}{\bf G}_0({\bf q}) {\bf T}; \ \ \ {\bf F}({\bf S})=\tilde {\bf T}{\bf F}({\bf q}) {\bf T},
\end{equation}
\textcolor{black}{where $\bf T$ is the symmetry transformation matrix determined by the molecular point group.} \textcolor{black}{The normalization of the ${\bf L}$ matrix is chosen such that it satisfies ${\bf L}^\dagger {\bf G}_0^{-1}({\bf S}) {\bf L} = {\bf M}^R$, where the elements fulfill $(\mathcal{G}_0^{-1})_{\iota \iota} = M^R_{\iota \iota}$. Here, $\mathcal{G}_0$ denotes the normal Wilson matrix evaluated at equilibrium, which defines the reduced mass matrix ${\bf M}^R$. Under this normalization, the ${\bf L}$ matrix becomes adimensional, and the normal coordinates acquire units of length. Consequently, the derivatives of the mean polarizability are expressed in the appropriate physical units.} The diagonalization (\ref{GF}) defines the  new curvilinear coordinates through relation ${\bf S}={\bf L}{\cal S}$, where both components $\tilde{\cal S}=({\cal S}_1,{\cal S}_3)$ span the irreducible representation $\Sigma^+$  of the group ${\cal C}_{\infty v}$. These coordinates are appropriate to describe the stretching degrees of freedom of asymmetric isotopologues of CO\textsubscript{2}. In the symmetric limit, where $m_{O_1}=m_{O_2}$ \textcolor{black}{is satisfied for the oxygen masses}, we find coincidence between ${\bf S}$ and ${\cal S}$, a property that justifies our selection for coordinates for the stretches \cite{marisol24}.

With the aim of establishing an algebraic realization of the nuclear motion,  the curvilinear coordinates are expanded in terms of stretching $\{{\cal Q}_\iota;\ \iota=1,3\}$ and bending $Q_\pm$ normal coordinates which, by definition, are rectilinear. The expansion up to cubic terms takes the form
\begin{subequations}
	\label{curv}
	\begin{eqnarray}
		\mathcal{S}_1&=&\mathcal{Q}_1+C_1\, \mathcal{Q}_1{Q}_+{Q}_-+C_2\, \mathcal{Q}_3{Q}_+{Q}_-+C_3\, {Q}_+{Q}_-, \\[0.15 cm]
		{S}_{+}&=&{Q}_{+}+ C_4\, {Q}_{+}\mathcal{Q}_1+C_5\, {Q}_{+}\mathcal{Q}_3+C_6\, {Q}_{+}\mathcal{Q}_1^2 \nonumber \\
		&+&C_7\, {Q}_{+}\mathcal{Q}_3^2+C_8\, {Q}_{+}\mathcal{Q}_1\mathcal{Q}_3-2\, C_9\,{Q}_{+}^2{Q}_{-}, \\[0.15 cm]
		{S}_{-}&=&{Q}_{-}+ C_4\, {Q}_{-}\mathcal{Q}_1+C_5\, {Q}_{-}\mathcal{Q}_3+C_6\, {Q}_{-}\mathcal{Q}_1^2 \nonumber \\
		&+&C_7\, {Q}_{-}\mathcal{Q}_3^2+C_8\, {Q}_{-}\mathcal{Q}_1\mathcal{Q}_3-2\, C_9\, {Q}_{+}{Q}_{-}^2, \\[0.15 cm]
		\mathcal{S}_3&=&\mathcal{Q}_3+C_{10}\, \mathcal{Q}_3{Q}_+{Q}_-+C_{11}\, \mathcal{Q}_1{Q}_+{Q}_-+C_{12}\, {Q}_+{Q}_-.
	\end{eqnarray}
\end{subequations}
with $\{C_i;\ i=1,\dots, 12\}$ being constants that depend on the structure constants and ${\bf L}$ matrix elements. \textcolor{black}{These coefficients can be calculated from the tensor elements ${\bf L}$ in Ref.~\cite{hoy}, requiring coordinates (\ref{curv}) to coincide with rectilinear normal coordinates for infinitesimal displacements.} The expressions for these constants are presented in the \ref{apa}. An algebraic realization of the expansions is obtained inserting the usual representation  of normal coordinates and  conjugate momenta in terms of bosonic  operators
\begin{subequations}
	\label{nalg}
\begin{eqnarray}
	\label{acal}\hat{\cal A}_\iota^\dagger&=&\frac{1}{\sqrt 2}\left(\sqrt{\frac{\Omega_\iota}{\hbar \mathcal{G}_{\iota\iota}}}{\cal Q}_\iota-i\sqrt{\frac{\mathcal{G}_{\iota\iota}}{\hbar \Omega_\iota}}{\cal P}_\iota\right), \\[0.25 cm]
	\label{tbend} \hat{\tau}_\pm^\dagger&=&\frac{1}{\sqrt 2}\left(\sqrt{\frac{\omega}{\hbar g_{aa}^{(0)}}}{Q}_\pm+i\sqrt{\frac{g_{aa}^{(0)}}{\hbar \omega}}{P_\mp}\right).
\end{eqnarray}
\end{subequations}
The frequencies of the stretching modes are related to the eigenvalues of Eq.~(\ref{GF}), $\Omega_\iota=\sqrt{\mathcal{F}_{\iota\iota}\mathcal{G}_{\iota\iota}}=(\Lambda_{\iota\iota})^{-1/2}$,  where  $\mathcal{F}_{\iota \iota}=({\bf L}^\dagger {\bf F}({\bf S}) {\bf L})_{\iota \iota}$, \textcolor{black}{and the frequencies of the degenerate bending modes are given by $\omega=\sqrt{f_{aa} g_{aa}^{(0)}}$, where the superscript index $(0)$ indicates that the element of the Wilson matrix elements $g^{(0)}_{aa}$   are evaluated at equilibrium geometry} \cite{Mar2018}. In the following section we introduce a canonical transformation between the normal and local bosonic operators. This transformation provides the foundation of a local/normal mode transition formalism,  allowing  the calculation of the matrix elements of the molecular polarizability in terms of the  anharmonic basis provided by the model defined in terms of the  dynamical algebra $su_1(2)\oplus su(3)\oplus su_2(2)$.

\section {Degree of normality}
\label{sec-localnormal}

The general formalism of the local/normal mode transition has been presented in Ref.~\cite{suarez1}. Here we discuss the salient features which allow us to define a parameter to quantify the degree of normality of the CO\textsubscript{2} isotopologues. The parameter is defined through the  analysis of normal modes starting in internal coordinates. It is well known the non-linear dependence of the local coordinates $\{q_i,\ i=1,2\}$ in terms of Cartesian displacement coordinates. The expansion up to linear terms defines the rectilinear internal coordinates $\{q_i',\ i=1,2\}$ with canonical conjugate momenta $\hat{p}_i'=-i\hbar\, \partial/\partial q_i'$. Introducing the definitions of the symmetry adapted coordinates in terms of the rectilinear local coordinates, the operator (\ref{acal}) takes the form
\begin{eqnarray}
	\hat{\cal A}^\dagger_\iota={1 \over 2}  \sum_\alpha \left\{ \sqrt{\frac{\Omega_\iota}{\hbar {\cal G}_{\iota\iota}}} ({\bf L}^{-1})_{\iota \alpha} (q_1'+(-1)^{(\alpha+1)}  q_2')-i\sqrt{\frac{{\cal G}_{\iota\iota}}{\hbar \Omega_\iota}}(\tilde{\bf L})_{\iota \alpha} (p_1'+(-1)^{(\alpha+1)}  p_2')\right\},
\end{eqnarray} 
where $\alpha=1$ for {\it gerade} (g) and $\alpha=2$ for the {\it ungerade} (u) stretching mode. We now introduce the bosonic ladder realization
\begin{equation}
	\label{pyM}
	q_j'={1 \over \sqrt{2}} \sqrt{{\hbar g_{jj}^{(0)} \over  \omega_j}} \left( \hat a_j^\dagger+\hat a_j \right); \qquad p_j'={i \over \sqrt{2}} \sqrt{\hbar \omega_j \over g_{jj}^{(0)}} \left( \hat a_j ^\dagger-\hat a_j  \right), 
\end{equation}
to obtain the expansion of the normal modes in terms of the local modes \cite{marisol24}
\begin{equation}
	\label{Aa}
	\hat{\cal A}^\dagger_\iota=  \sum_j(c^{( \iota,j)}_+ \hat a^\dagger_j+c^{(\iota, j)}_-\hat a_j),
\end{equation}
with coefficients
\begin{eqnarray}
	\label{cescof}
	c^{(\iota,j)}_\pm={1 \over 2 \sqrt{2}}  \sum_\alpha (-1)^{(\alpha+1) (1-\delta_{j1})}\left[ \sqrt{\frac{\Omega_\iota\, g_{jj}^{(0)}}{\omega_j\, {\cal G}_{\iota\iota}}}({\bf L}^{-1})_{\iota \alpha}\pm \sqrt{\frac{\omega_j\, {\cal G}_{\iota\iota}}{\Omega_\iota\, g_{jj}^{(0)}}} (\tilde{\bf L})_{\iota \alpha} \right],
\end{eqnarray}
\textcolor{black}{where $\omega_j$ and $g_{jj}^{(0)}$, with $j=1,2$, denote the frequencies and the corresponding Wilson matrix elements for the local modes, respectively.} We should notice that the commutator $[\hat{\cal A}_\iota,\hat{\cal A}^\dagger_{\iota'}]=\delta_{\iota\iota'}$ imposes the condition 
\begin{equation}
	\sum_{j=1}^2(|c_+^{(\iota,j)}|^2-|c_-^{(\iota,j)}|^2)=1,
\end{equation}
which is indeed satisfied.
For symmetrical isotopologues the 
 expression (\ref{cescof}) for the coefficients simplifies to 
\begin{eqnarray}
	\label{cpmS}
	c^{(\alpha,j)}_\pm={1 \over 2 \sqrt{2}}    (-1)^{(\alpha+1) (1-\delta_{j1})} \bigg({1+(-1)^{(\alpha+1)} x_f \over 1+(-1)^{(\alpha+1)} x_g} \bigg)^{1/4}\bigg[ 1 \pm   \bigg( {1+(-1)^{(\alpha+1)} x_g \over 1+(-1)^{(\alpha+1)} x_f} \bigg)^{1/2} \bigg],
\end{eqnarray}
with the definitions $x_f=f_{rr'}/f_{rr}$ and $x_g=g_{rr'}^{(0)}/g_{rr}^{(0)}$\textcolor{black}{, where $f_{ij}$ denote the local-mode force constants}.

The important property  of the expansion (\ref{Aa}) is that, in the pure local limit, the annihilation contributions $c_-^{(\iota,j)}$ vanish and, consequently, the parameter
\begin{equation}
	\label{DeltaM}
	\delta_-={1 \over 2}\sum_{\iota=1}^2 \sum_{j=1}^2 |c^{(\iota,j)}_-|^2,
\end{equation}
measures the degree of normality of the system. In Tables \ref{ces1} and \ref{ces1S} the parameter $\delta_-$  is displayed for the stable asymmetric and symmetric isotopologues, respectively. For this calculation we use the force constants from Ref. \cite{marisol24} assuming the Born-Oppenheimer approximation. We observe that the main contributions to the operators $\hat{\cal A}^\dagger_\iota$ correspond to the local creation operators $\hat{a}_j^\dagger$ with different weights. For symmetric isotopologues it holds that $|c_\pm^{(\alpha,1)}|=|c_\pm^{(\alpha,2)}|$, a property that reflects the equivalence of local modes into the normal modes. In the case of  asymmetric isotopologues such symmetry  dependence does not exist because the lack of inversion center. For the isotopologue series of CO\textsubscript{2}, the degree of normality ranges from 0.0338 to 0.0418, which represents a remarkably large value compared with that of local-mode molecules such as H\textsubscript{2}O, whose parameter is $\delta_- = 5.47\times10^{-7}$ \cite{suarez1}. \textcolor{black}{This parameter represents an alternative to the Child and Halonen's parameter used to measure the degree of locality \cite{lawton,child84,halonen98,child2}.}

\begin{table}[h!]
\centering
\caption{Coefficients $c_\pm^{(\iota,j)}$ involved in the expansion (\ref{Aa}) for the asymmetric isotopologues. The order corresponds to ascending degree of normality.}
\label{ces1}

\scriptsize
\begin{tabular}{cccccccc}
\toprule
Isotopologue & $\iota$-th mode & $c_+^{(\iota,1)}$ & $c_+^{(\iota,2)}$ & $c_-^{(\iota,1)}$ & $c_-^{(\iota,2)}$ & $\sum_j |c^{(\iota,j)}_-|^2$ & $\delta_-$ \\
\midrule
(637) & 1 & 0.7156 & 0.7346 & 0.1619 & 0.1599 & 0.0518 & 0.0349 \\
 & 3 & 0.7230 & -0.7037 & -0.0930 & 0.0968 & 0.0180 & \\
(638) & 1 & 0.7077 & 0.7435 & 0.1657 & 0.1617 & 0.0536 & 0.0360 \\
 & 3 & 0.7315 & -0.6951 & -0.0920 & 0.0993 & 0.0183 & \\
(738) & 1 & 0.7181 & 0.7349 & 0.1679 & 0.1660 & 0.0557 & 0.0372 \\
 & 3 & 0.7222 & -0.7051 & -0.0948 & 0.0983 & 0.0186 & \\
(627) & 1 & 0.7183 & 0.7357 & 0.1701 & 0.1682 & 0.0572 & 0.0381 \\
 & 3 & 0.7225 & -0.7048 & -0.0953 & 0.0990 & 0.0189 & \\
(628) & 1 & 0.7112 & 0.7439 & 0.1739 & 0.1703 & 0.0592 & 0.0392 \\
 & 3 & 0.7303 & -0.6970 & -0.0944 & 0.1014 & 0.0192 & \\
(728) & 1 & 0.7208 & 0.7362 & 0.1763 & 0.1746 & 0.0616 & 0.0405 \\
 & 3 & 0.7217 & -0.7061 & -0.0971 & 0.1004 & 0.0195 & \\
\bottomrule
\end{tabular}
\end{table}

\begin{table}[h!]
\centering
\caption{Coefficients $c_\pm^{(\alpha,j)}$ involved in the expansion (\ref{Aa}) for the symmetric isotopologues. The order corresponds to ascending degree of normality.}
\label{ces1S}

\scriptsize
\begin{tabular}{cccccccc}
\toprule
Isotopologue & $\alpha$-th mode & $c_+^{(\alpha,1)}$ & $c_+^{(\alpha,2)}$ & $c_-^{(\alpha,1)}$ & $c_-^{(\alpha,2)}$ & $\sum_j |c^{(\alpha,j)}_-|^2$ & $\delta_-$ \\
\midrule
(636) & g & 0.7245 & 0.7245 & 0.1579 & 0.1579 & 0.0499 & 0.0338 \\
 & u & 0.7133 & -0.7133 & -0.0941 & 0.0941 & 0.0177 & \\
(737) & g & 0.7259 & 0.7259 & 0.1640 & 0.1640 & 0.0538 & 0.0361 \\
 & u & 0.7136 & -0.7137 & -0.0958 & 0.0958 & 0.0183 & \\
(626) & g & 0.7263 & 0.7263 & 0.1660 & 0.1660 & 0.0551 & 0.0368 \\
 & u & 0.7136 & -0.7136 & -0.0963 & 0.0963 & 0.0189 & \\
(838) & g & 0.7273 & 0.7273 & 0.1700 & 0.1700 & 0.0578 & 0.0384 \\
 & u & 0.7138 & -0.7138 & -0.0974 & 0.0974 & 0.0190 & \\
(727) & g & 0.7278 & 0.7278 & 0.1724 & 0.1724 & 0.0595 & 0.0393 \\
 & u & 0.7139 & -0.7139 & -0.0980 & 0.0980 & 0.0192 & \\
(828) & g & 0.7293 & 0.7293 & 0.1786 & 0.1786 & 0.0638 & 0.0418 \\
 & u & 0.7141 & -0.7141 & -0.0995 & 0.0995 & 0.0198 & \\
\bottomrule
\end{tabular}
\end{table}

Finally, it is worth emphasizing the physical meaning of the degree of normality. We observe that, in the local limit,
\begin{equation}
    \lim_{\delta_-\to0}\hat{P}_N=\hat{P}_L,
\end{equation}
where $\hat{P}_N$ and $\hat{P}_L$ denote the conserved quantities referred to as the normal and local polyad numbers, respectively \cite{marisolpoly2}. In this context, the transition toward a normal-mode description gives rise to a  polyad breaking,  which is equivalent to the loss of an approximate constant of motion, a phenomenon of fundamental significance in contemporary physical theories. \textcolor{black}{To construct a vibrational Hamiltonian that preserves a local polyad number, we introduce the transformation between normal and local operators 
\begin{equation}
    \label{cA}\hat{\mathcal{A}}_\iota^\dagger=\sum_{j=1,2} c_+^{(\iota,j)}\hat{a}_j^\dagger,
\end{equation}
which must be subsequently renormalized to ensure that the canonical bosonic commutation relations are fulfilled, i.e., $[\hat{\cal A}_\iota,\hat{\cal A}_{\iota'}^\dagger]=\delta_{\iota\iota'}$ as shown in Refs.~\cite{marisol24,suarezfcn,suarezocs}. Equation~(\ref{cA}) coincides with the transformations defined in Refs.~\cite{halonen98,valle}.}

\section{Vibrational description}
\label{sec-vibdesc}

\textcolor{black}{A comprehensive vibrational description of the isotopologue series was reported in Refs.~\cite{marisol24,marisol22} using the $SU_1(2)\times SU(3)\times SU_2(2)$ anharmonic model. Within this formalism, the effective Hamiltonian conserving the polyad $P=2\nu_1+\nu_2+2\nu_2$ is chosen as
\begin{align}
\hat{H}&= \sum_{l=1,3} \frac{\hbar \Omega_l}{2}
\left(\hat{\mathcal{A}}_\iota^\dagger \hat{\mathcal{A}}_\iota
+ \hat{\mathcal{A}}_\iota \hat{\mathcal{A}}_\iota^\dagger\right) + \frac{\hbar\omega}{2}
\sum_{\alpha=+,-}
\left(\hat{\tau}_\alpha^\dagger \hat{\tau}_\alpha
+ \hat{\tau}_\alpha \hat{\tau}_\alpha^\dagger\right) \nonumber\\
&\quad + \sum_{i,j} x_{ij}
\left\{\hat{\nu}_i,\hat{\nu}_j\right\}
+ g_{22} \hat{l}^{\,2}
+ \alpha_F \hat{F} + \alpha_F^{(1)} \left\{\hat{\nu}_1,\hat{F}\right\}\nonumber \\
&\quad+ \alpha_F^{(3)} \left\{\hat{\nu}_3,\hat{F}\right\}
+ d_N \hat{D}
\label{vibham}
\end{align}
with definitions
\begin{subequations}
    \begin{eqnarray}
        \hat{\nu}_\iota&=&\hat{\cal A}_\iota^\dagger\hat{\cal A}_\iota; \qquad \iota=1,3\\
        \hat{\nu}_2&=&\hat{\tau}_+^\dagger \hat{\tau}_+ +\hat{\tau}_-^\dagger \hat{\tau}_-\\
        \hat{\ell}&=&\hat{\tau}_+^\dagger \hat{\tau}_+ -\hat{\tau}_-^\dagger \hat{\tau}_-\\
        \hat{F}&=&\hat{\cal A}_1^\dagger\hat{b}_+\hat{b}_-+\mathrm{H.c.} \label{Fermi}\\
        \hat{D}&=&\hat{\cal A}_1^\dagger\hat{\cal A}_1^\dagger\hat{\cal A}_3\hat{\cal A}_3+\mathrm{H.c.}
    \end{eqnarray}
\end{subequations}
The Fermi interaction~(\ref{Fermi}) constitutes a key contribution to the vibrational description. Within the framework of the present study, the Fermi interaction strength $\alpha_F$ is found to correlate with $\delta_-$: specifically, it increases with the degree of normality, as evidenced by a larger splitting of the stretching modes~\cite{marisol24,marisol22}.}

\textcolor{black}{We have assumed the transformation (\ref{cA}) applying the mapping
$\hat{a}_i(\hat{a}^\dagger_i)\to\hat{b}_i(\hat{b}^\dagger_i)$ for the stretching modes with $i=1,2$ and $\hat{\tau}_\pm(\hat{\tau}^\dagger_\pm)\to\hat{b}_\pm(\hat{b}^\dagger_\pm)$ for the bending degrees of freedom, which are defined through the matrix elements \cite{marisol24}
\begin{subequations}
    \label{MEs}
    \begin{eqnarray}
    \hat b^\dagger_i |\Psi_{v_i}^{j} \rangle&=&\sqrt{(v_i+1)(1-(v_i+1)/\kappa_i)}
    ~| \Psi_{v_i+1}^{j} \rangle, \\[0.2 cm] 
    \hat b_i|\Psi_{v_i}^{j}\rangle&=&\sqrt{v_i(1-v_i/\kappa_i)} ~|\Psi_{v_i-1}^{j} \rangle, \\[0.2 cm]
    \hat b^\dagger_\pm|[N];n^{\ell}\rangle&=&\sqrt{\left(\frac{n\pm{\ell}}{2}+1\right)\left(1-\frac{n}{N}\right)}\ |[N];(n+1)^{{\ell}\pm 1} \rangle,\\
    \hat b_\pm |[N];n^{\ell}  \rangle&=&\sqrt{\left( \frac{n \pm {\ell}}{2} \right) \left(1-\frac{n-1}{N} \right)} \  |[N];(n-1)^{{\ell} \mp 1}\rangle,
    \end{eqnarray}
\end{subequations}
where $\kappa_i=2j_i+1$ with $v_i=1,2,\dots,j-1$. It should be emphasized that the basis functions and their operators depend on three external parameters, $\{N, j_1, j_2\}$, which are directly related to the anharmonicity and the dimensionality of the associated potential.}

\textcolor{black}{The eigenstates conserving polyad $P$, denoted as $|\nu,P \rangle$, are expressed as linear combinations of the local basis
\begin{equation}
    |P;j_1,j_2,[N];v_1 v_2 n^\ell\rangle=|j_1v_1 \rangle \otimes |j_2 v_2 \rangle \otimes |[N];n_+ n_- \rangle
\end{equation} 
projected to the irreducible representation $\Gamma(\pm\ell)$:
\begin{equation}
\label{base}
|\nu,\Gamma(\pm\ell),P \rangle=\sum_{v_1,v_2, n_+,n_-} C^{(\Gamma,P)}_{v_1,v_2, n_+,n_-}|P;j_1,j_2,[N];v_1 v_2 n_+ n_-\rangle,
\end{equation}  
where $|j_i v_i \rangle$ are eigenstates associated with the bound Morse states with $v_i=0,1,2,\dots,j_i-1$, while $|[N];n^{\pm \ell} \rangle=|[N];n_+,n_- \rangle$ are the bending states associated with the $SU(3)$ model, with vibrational angular momentum $\ell=n_+-n_-$, bending quantum number $n=n_++n_-$ and $N$ as a parameter that specifies the dimension of the Hilbert space to describe the bending modes. The vibrational labels are given in accordance with  the subgroup chain $\mathcal{D}_{\infty h} \supset \mathcal{C}_{\infty v}$, associated with angular momentum projection (${\cal C}_{\infty v}\equiv O(2)$), which explain the notation $\Gamma(\pm \ell)$ for the corresponding irreps.}

\textcolor{black}{The eigenstates (\ref{base}) can also be expressed in terms of a normal mode basis $|j_1,j_2,[N];\nu_1\nu_2^\ell\nu_3\rangle$ in the following form
\begin{equation}
    |j_1,j_2,[N];\nu_1\nu_2^\ell\nu_3\rangle=\sum_{v_1,v_2,n_+,n_-}A^{(\nu_1,\nu_2,\nu_3)}_{v_1,v_2,n_+,n_-}|j_1,j_2,[N];v_1 v_2 n_+ n_-\rangle.
\end{equation}
In our method, a normal basis can be constructed by diagonalizing the normal number operators $\{\hat{\nu}_\iota;\ \iota=1,2,3\}$ within a harmonic local basis and then using the same coefficients to build a normal mode basis expressed in the local anharmonic basis. Hence, the eigenkets can be recast in terms of the normal basis
\begin{equation}
\label{normal}
|\nu,\Gamma(\pm\ell),P \rangle=\sum_{v_1,v_2, n_+,n_-} B^{(\Gamma,P)}_{\nu_1,\nu_2,\nu_3}|P;j_1,j_2,[N];\nu_1\nu_2^\ell\nu_3\rangle,
\end{equation}
with the relation between the coefficients
\begin{equation}
    C^{(\Gamma,P)}_{v_1,v_2, n_+,n_-}=\sum_{\nu_1,\nu_2\nu_3} B^{(\Gamma,P)}_{\nu_1,\nu_2,\nu_3}\, A^{(\nu_1,\nu_2,\nu_3)}_{v_1,v_2,n_+,n_-}.
\end{equation}
The labeling of the eigenstates is given according to the major contribution in (\ref{normal}).}

\section {Model for Raman spectra simulation} 
\label{sec-raman}

The present model can be considered as the generalization  of the previously proposed scheme, devised to simulate the Raman spectrum of the principal CO\textsubscript{2} isotopologue \cite{mariano,marisol2020}, to the asymmetric triatomic linear molecules. \textcolor{black}{In the present work, we restrict the simulations to pure vibrational transitions within the rigid-rotor approximation.
} More accurate simulations are available  taking into account  both isotropic and  anisotropic scattering including all ro-vibrational transitions for some of the stable isotopologues \cite{lill1,lill2}. \textcolor{black}{It should be emphasized that the standard approach for computing spectral intensities is based on the double harmonic approximation, in which the potential energy surface (PES) is truncated at second order and the polarizability surface is restricted to linear terms \cite{Barone2014}. Beyond this approximation, vibrational perturbation theory and contact transformation methods have been widely employed to incorporate anharmonic effects, typically within a harmonic-oscillator basis set \cite{Bloino2015,Krasnoshchekov2010}. The present approach goes beyond the harmonic approximation by employing anharmonic vibrational wave functions obtained from a higher-order PES expansion \cite{marisol24,marisol22},  evaluating the Raman transition moments in the framework of an interacting anharmonic oscillator model, as we next explain.
}

\textcolor{black}{The Raman line intensities for the isotropic contribution} are obtained with the trace of the differential cross section $[\mathrm{m^2/sr}]$, which for a gas sample at thermal equilibrium is given by \cite{tejeda,alvarez}
\begin{equation}
    \label{RTI}
    \bigg( {d \sigma(\bar\nu) \over d\Omega} \bigg)^{{\rm trace}}_{i \to f}=\bigg({\pi \over \epsilon_0} \bigg)^2 {(\bar\nu_0-\Delta\bar\nu)^4 \over Z_{\rm vib}(T)}\,g_{\rm ns}\, g_{\nu_i}\, |M_{if}|^2 \exp\bigg({-h c \bar\nu_i \over k_B T}\bigg), 
\end{equation}
where $\epsilon_0$ is the vacuum permittivity, $\bar\nu_0$ $\mathrm{[cm^{-1}]}$ is the wave number of the exciting radiation, \textcolor{black}{the Raman shift is $\Delta\bar\nu=\bar\nu_f-\bar\nu_i$, with $\bar\nu_i$ and $\bar\nu_f$ representing the vibrational term values for the initial and final states (in cm$^{-1}$)}. The factors $g_{\mathrm{ns}}$ and $g_{\nu_i}$ represent the nuclear spin and vibrational degeneracies of the initial state, and the partition function at temperature $T\ \mathrm{[K]}$ is computed as $Z_{\rm vib}(T)=\sum_j g_{\nu_j}\, e^{-hc{\bar \nu}_j/k_BT}$, where $g_{\nu_j}$ is the degeneracy of the state with energy $hc\bar\nu_j$. A set of values of the vibrational partition function computed for the main isotopologue of CO$_2$ are given in Ref.~\cite{marisol2020}. Because of the selection of the trace in Eq.~(\ref{RTI}), the contribution of the intensities  are due to the matrix elements $M_{ij}=\langle \nu_i|\bar \alpha| \nu_j \rangle$  \textcolor{black}{$\mathrm{[C\cdot m^2/V]}$} of the totally symmetric mean polarizability ${\alpha}_{\Sigma^+}\equiv\bar \alpha$. The expression (\ref{RTI}) involves  pure vibrational transitions without  any reference to the rotational degrees of freedom. \textcolor{black}{Its justification, based on the rigid-rotor approximation, is presented in the Supplementary Material and also analyzed in Ref.~\cite{EduardoRaman2026}.} 
%

\textcolor{black}{In order to calculate the matrix elements} $M_{ij}$ in Eq.~(\ref{RTI}) beyond the harmonic approximation, it is necessary to obtain an expression of the mean polarizability in terms of the generators of the dynamical group. To achieve this goal the polarizability is first expanded around the molecular equilibrium configuration in terms of the curvilinear symmetry coordinates ${\cal S}$. Up to cubic terms we have
\begin{eqnarray}
	\label{alpha}
    \bar
\alpha &=& \bar \alpha_0+\sum_{\iota=1,3} \bigg( {\partial \bar \alpha \over \partial{\cal S}_\iota}\bigg)_0 {\cal S}_\iota +{1 \over 2}\sum_{\iota=1,3} \bigg( {\partial^2 \bar \alpha \over \partial {\cal S}_\iota^2} \bigg)_0 {\cal S}_\iota^2 \nonumber \\
    &+&\bigg({\partial^2 \bar \alpha \over \partial {\cal S}_1 \partial {\cal S}_3} \bigg)_0 {\cal S}_1 {\cal S}_3+\bigg( {\partial^2 \bar \alpha \over \partial{S}_+\partial S_-}  \bigg)_0 S_+S_-\nonumber\\
    &+&\sum_{\iota=1,3} \bigg({ \partial^3 \bar \alpha \over  \partial {\cal S}_\iota \partial{S}_+\partial S_-}  \bigg)_0 {\cal S}_\iota{S}_+S_-.
\end{eqnarray}
For symmetric isotopologues the matrix  ${\bf G}_0({\bf S}){\bf F}({\bf S})$ is diagonal 
 and, consequently, the stretching derivatives in Eq.~(\ref{alpha}) simplify to $\partial/\partial {\cal S}_1=\partial/\partial S_g$ and $\partial/\partial {\cal S}_3=\partial/\partial S_u$. We now return to the expansions (\ref{curv}), which can be realized in algebraic form by introducing  the bosonic expressions (\ref{nalg}).  The obtained expansions, when substituted into the  polarizability (\ref{alpha}), leads to its algebraic representation  in terms of the stretching $\hat{\cal A}_{\iota}^\dagger(\hat{\cal A}_\iota)$ and bending $\hat{\tau}_\pm^\dagger(\hat{\tau}_\pm)$ bosonic normal operators. If we considered this expansion, the matrix elements would be calculated with the harmonic limit of the basis (\ref{base}). \textcolor{black}{In order to be consistent with the anharmonic basis in terms of Morse and $SU(3)$ states, we insert the canonical transformation (\ref{cA}) applying the mapping
$\hat{a}_i(\hat{a}^\dagger_i)\to\hat{b}_i(\hat{b}^\dagger_i)$ for the stretches:
\textcolor{black}{\begin{equation}
\hat{\mathcal{A}}_\iota^\dagger=\sum_{j=1,2} c_+^{(\iota,j)}\hat{b}_j^\dagger,
\end{equation}}
and $\hat{\tau}_\pm(\hat{\tau}^\dagger_\pm)\to\hat{b}_\pm(\hat{b}^\dagger_\pm)$ for the bends. Finally, we note that the mean polarizability is now in terms of the generators of the dynamical algebra together with the potential parameters:}
\begin{equation}
	\hat{\alpha}_{\Sigma^+}=\bar{\alpha}(\kappa_i,[N];\hat{b}_i^\dagger,\hat{b}_i,\hat{b}_\pm^\dagger,\hat{b}_\pm)\ ; \qquad i=1,2.
\end{equation}

With help of the matrix elements (\ref{MEs}) we are able to calculate the transition moments $M_{ij}$ required to simulate the trace of the differential cross section (\ref{RTI}).
However, the calculation of the matrix elements $M_{ij}$ involves the derivatives of the polarizability. Their estimation is obtained by fitting $42$ experimental transition moments,  
reported for the main isotopologue of CO$_2$~\cite{tejeda,alvarez,chrysos},
using the wave functions obtained \textcolor{black}{through the method  presented in Refs. \cite{marisol24,marisol17,marisol22}}. This procedure is carried out by  minimizing a root mean square deviation between experimental $|M^{({\rm exp})}|$ and calculated $|M^{({\rm calc})}|$ transition moments for the principal isotopologue, with the following definition for the root mean square deviation
\begin{equation}
	\label{rms}
	\mathrm{rms}=\sqrt{\frac{1}{\cal N}\sum_\beta s_\beta\left( \log|M_\beta^{({\rm exp})}|-\log|M_\beta^{({\rm calc})}| \right)^2},
\end{equation}
with a normalization ${\cal N}=\sum_\beta s_\beta$, given in terms of the weight factors $s_\beta=1/\sigma_\beta$, to take into account the experimental uncertainty $\sigma_\beta$ of the transition moments.  The results of the fit are displayed in \textcolor{black}{the Supplementary Material}. The derivatives of mean polarizability of the main isotopologue of CO$_2$ were fitted with a resulting deviation of rms $=0.061$. The obtained derivatives can be used to simulate the spectra of any isotopologue.

For symmetric isotopologues,  the  derivatives of the polarizability \textcolor{black}{with respect curvilinear coordinates (\ref{qt})} are mass independent, a consequence of the selected \textcolor{black}{set} $S_{g,u}$.  In contrast, the derivatives of the polarizability with respect to the coordinates ${\cal S}_\iota$ are mass dependent, and they can be obtained  employing the chain rule involving  the derivatives with respect to the curvilinear coordinates ${\cal S}$ and ${\bf S}$:
\begin{equation}
	\label{chain}
	\frac{\partial}{\partial {\cal S_\iota}}=\sum_{\alpha=1}^2 {\bf L}_{\alpha\iota}\frac{\partial}{\partial S_\alpha},
\end{equation}
with the correspondence $\iota=1\to1$ and $\iota=3\to2$. The expression (\ref{chain}) allows us to obtain mass dependent derivatives in terms of the mass independent coordinates considered for all the isotopologues in  accordance with the Born-Oppenheimer approximation. Table \ref{polsur} shows our predictions for polarizability surface of the isotopologues series. These derivatives will be used to simulate the Raman intensities for some isotopologues whose wave functions are available using the  effective Hamiltonian energy fit described in Refs.~\cite{marisol24,marisol17,marisol22}. In addition, the mass dependent derivatives of the polarizability allows us to explore a correlation with the parameters displayed in Table \ref{ces1}. Furthermore, {\it ab initio} methodologies have been employed in the description of the electrical properties of carbon dioxide. The first-order and certain second-order derivatives for symmetric and antisymmetric stretches were computed utilizing post-Hartree-Fock approaches \cite{Maroulis2003,Haskopoulos2006,Beil2021}. In the following section we present a description of the experiment in which we measure the Raman spectra of the complete series of stable isotopologues. These measurements will allow us to validate our theoretical scheme to simulate the Raman intensities and establish the  correlations in accordance with the molecular normal mode degree.

\begin{table}[h!]
\centering
\caption{\textcolor{black}{Mean polarizability derivatives of the CO\textsubscript{2} isotopologues calculated using the chain rule (\ref{chain}) in the fitted derivatives of principal isotopologue (626). For all symmetric isotopologues, the derivatives are identical to the fitted values for the main isotopologue}.}
\label{polsur}

\textcolor{black}{\scriptsize
\setlength{\tabcolsep}{4pt}
\begin{tabular}{lccccccc}
\toprule
Derivative (units) & Symmetric\textsuperscript{a} & (637) & (638) & (738)  & (627) & (628) & (728)  \\
\midrule
$\displaystyle\left(\frac{\partial\bar{\alpha}}{\partial\mathcal{S}_1}\right)_0$ ($10^{-30}$ CV$^{-1}$m) & 3.219 & 209.24 & 111.08 & 236.48 & 228.09 & 121.05 & 257.69 \\[0.2cm]
$\displaystyle\left(\frac{\partial\bar{\alpha}}{\partial\mathcal{S}_3}\right)_0$ ($10^{-30}$ CV$^{-1}$m) & \textemdash & 0.0416 & 0.0783 & 0.0368 & 0.0381 & 0.0719 & 0.0338 \\[0.2cm]
$\displaystyle\left(\frac{\partial^2\bar{\alpha}}{\partial\mathcal{S}_1^2}\right)_0$ ($10^{-20}$ CV$^{-1}$) & 2.559 & 10\,813.2 & 3047.42 & 13\,812.0 & 12\,849.0 & 3619.2 & 16\,400.0 \\[0.2cm]
$\displaystyle\left(\frac{\partial^2\bar{\alpha}}{\partial\mathcal{S}_3^2}\right)_0$ ($10^{-20}$ CV$^{-1}$) & 0.4664 & 0.4668 & 0.4679 & 0.4667 & 0.4676 & 0.4676 & 0.4667 \\[0.2cm]
$\displaystyle\left(\frac{\partial^2\bar{\alpha}}{\partial\mathcal{S}_1\partial\mathcal{S}_3}\right)_0$ ($10^{-20}$ CV$^{-1}$) & \textemdash & \multicolumn{6}{c}{1.683\textsuperscript{b}} \\[0.2cm]
$\displaystyle\left(\frac{\partial^2\bar{\alpha}}{\partial S_+\partial S_-}\right)_0$ ($10^{-20}$ CV$^{-1}$) & \multicolumn{7}{c}{0.2310\textsuperscript{c}} \\[0.2cm]
$\displaystyle\left(\frac{\partial^3\bar{\alpha}}{\partial\mathcal{S}_1\partial S_+\partial S_-}\right)_0$ ($10^{-10}$ CV$^{-1}$m$^{-1}$) & 1.767 & 114.84 & 60.96 & 66.44 & 125.19 & 129.80 & 141.43 \\[0.2cm]
$\displaystyle\left(\frac{\partial^3\bar{\alpha}}{\partial\mathcal{S}_3\partial S_+\partial S_-}\right)_0$ ($10^{-10}$ CV$^{-1}$m$^{-1}$) & \textemdash & 0.0228 & 0.0430 & 0.0202 & 0.0209 & 0.0395 & 0.0185 \\
\bottomrule
\end{tabular}
}
\begin{flushleft}
\scriptsize
\textsuperscript{a} Fitted derivatives of the (626) isotopologue. \textcolor{black}{This set is isotopically invariant.}\\
\textsuperscript{b} Approximate equal for all asymmetric species.\\
\textsuperscript{c} Same numerical value for the whole series.
\end{flushleft}
\end{table}

\section{Experimental measurements}
\label{sec-exp}

To produce samples with strong enrichment of selected carbon dioxide isotopologues several mixtures of gaseous carbon dioxide, either $^{12}$C or $^{13}$C with natural abundance of O isotopes, and water with an oxygen isotope label, either $^{16}$O, $^{17}$O or $^{18}$O, have been prepared in sealed quartz glass vials. During dissolution of the carbon dioxide into the water and its dissociation, rapid exchange and equilibration of oxygen isotopes was achieved resulting in exsolving carbon dioxide carrying the O label from the water. The sealed quartz glass vials with the isotopologues in the gas phase have been analyzed in a special holder with Neon light penetrating the quartz glass tube from below with a confocal LabRAM HR800 Evolution Horiba Jobin-Yvon Raman spectrometer. For the analyses a green Coherent VERDI laser with 532 nm wavelength and 2 W output power was used as light source. The laser was focused into the center of the quartz glass vial by an 10x objective connected to an upright Olympus BXFM microscope. Spectral resolution was achieved by using a grating with 2400 groves/mm. Sensitive detection was possible with a back-illuminated Symphony CCD with 2048$\times$512 pixels cooled by liquid nitrogen to $-$133~°C. All analyses were performed for 500 s integration time with 3 accumulations. The spectra were acquired as intensity in counts/s versus absolute wavelength in nm. The Neon line at 576.44188 nm was used as reference line, and the spectrum shifted to yield exact matching with the tabulated NIST data from Ref.~\cite{saloman}. The Raman shift was calculated using the exact laser wavelength of 532.131~nm. Data analyses, i.e., peak fitting and integration, was performed with the Horiba LabSpec software version 6.7.2. The results of the experimental intensities are displayed in \textcolor{black}{the Supplementary Material}. Further details on the methods can be found in  Ref.~\cite{christian}.

\section{Results and correlations}
\label{sec-results}

\begin{figure*}[htbp]
    \includegraphics[width=\textwidth]{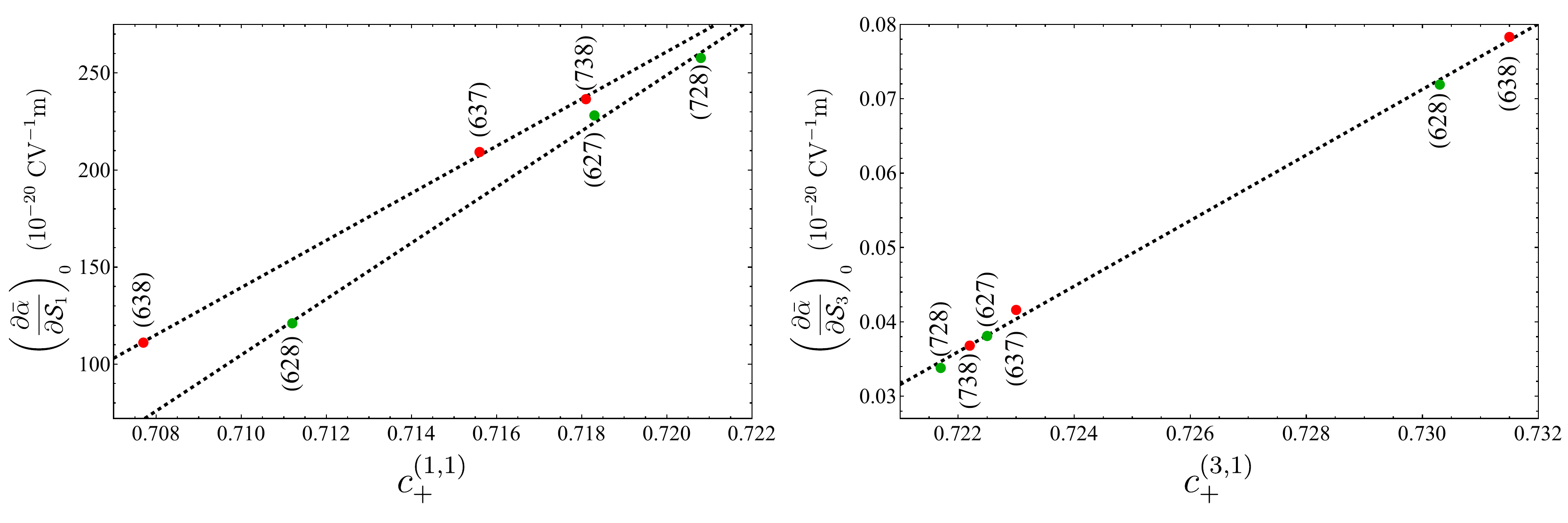}
	\caption{Correlation between parameters $c_\pm^{(\iota,j)}$ defined in Eq.~(\ref{cescof}) and the first order derivatives of the mean polarizability for the asymmetric isotopologues displayed in Table~\ref{polsur}. \textcolor{black}{Left and right panels correspond to the symmetric and asymmetric stretching modes, respectively.}}
    \label{dercor}
\end{figure*}

Using our theoretical model we have estimated the derivatives of the polarizability displayed in Table \ref{polsur}. In addition, we have presented an approach to measure the degree of normality $\delta_-$ of any molecule, but also the contribution of the local components $c_{\pm}^{(\iota,j)}$. We have found out a correlation for asymmetric isotopologues between the first order derivatives $(\partial \bar \alpha/\partial {\cal S}_1)_0$ and the coefficients $c_-^{(1,1)}$ as well as between  the derivatives  $(\partial \bar \alpha/\partial {\cal S}_3)_0$ and the coefficients $c_+^{(3,1)}$, both displayed in Figure \ref{dercor}. In the left display, the correlation is split in two sets, characterized by the mass of the carbon atom. We should mention that this feature also appears in the correlation between  the parameter $\delta_-$ and the Fermi interaction \cite{marisol24}.
In contrast, in the right display, a clear correlation appears for the full set of isotopologues.

The consequence of taking the most abundant molecule $^{12}$C$^{16}$O$_2$ as a reference to study the different isotopologues is manifested in the description of the PES using variational methods. In Ref. \cite{cerezo} the PES for  $^{12}$C$^{16}$O$_2$ is refined based on 49 pure vibrational frequencies (with $\ell=0$) using a variational method with a Morse-cosine expansion for the potential. This potential surface was used to provide a prediction for the rovibrational energies of the isotopologues. In particular, since in this work we are interested in the vibrational energies involved in the principal Raman transitions at room temperature, we compare the residuals of the corresponding transition energies of the set of isotopologues calculated variationally~\cite{cerezo} and \textcolor{black}{with an effective Hamiltonian~\cite{marisol24,marisol17,marisol22}} with respect to the experimental values. With this aim we define the Raman frequency shift errors for each symmetry, $\Sigma_g^+$ and $\Pi_u$ for the symmetric species and $\Sigma^+$ and $\Pi$ for the asymmetric ones:
\begin{equation}
\label{error}
    \text{error}=(\Delta\bar\nu_{+}^{({\rm exp})}-\Delta\bar\nu_{+}^{({\rm calc})})-(\Delta\bar\nu_{-}^{({\rm exp})}-\Delta\bar\nu_{-}^{({\rm calc})}),
\end{equation}
where $\nu_\pm$ labels the Fermi dyad involved in the principal Raman transitions at room temperature. 

\begin{figure*}[htbp]
	\includegraphics[width=\textwidth]{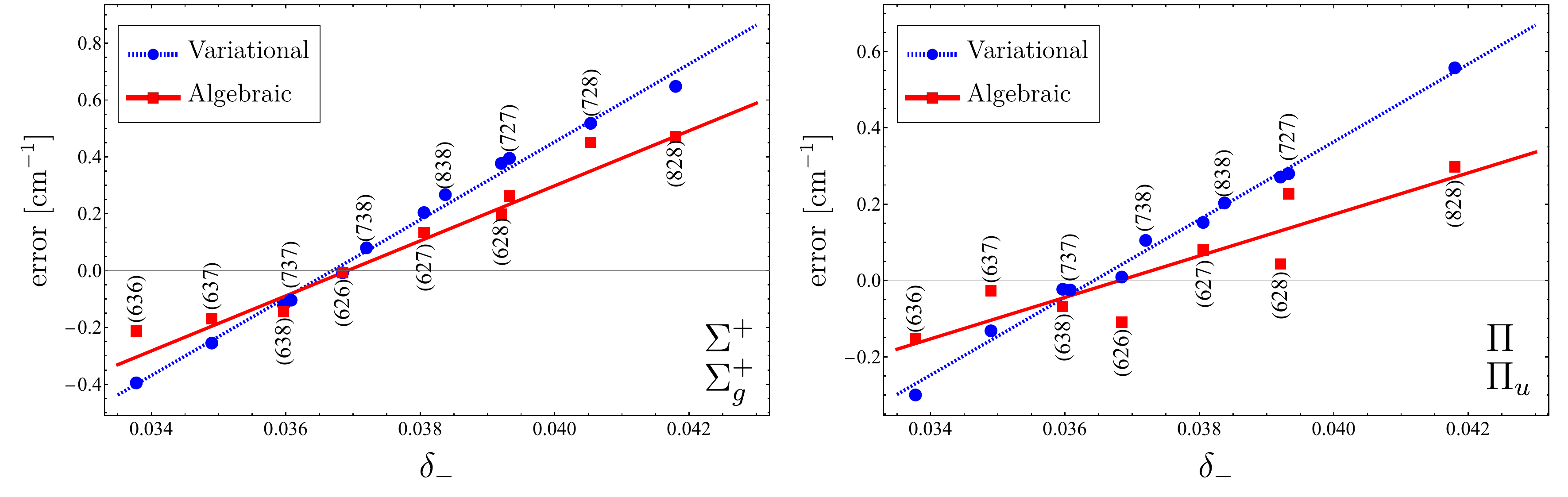}
	\caption{Correlation between the parameter $\delta_-$ defined in Eq.~(\ref{DeltaM}) and the error in the predicted \textcolor{black}{Raman} shifts associated with $\nu_+$ and $\nu_-$ dyads \textcolor{black}{(notation defined in Tables~\ref{symshift} and \ref{asymshift})}. Two methods are compared: a variational approach from {Cerezo et al.}~\cite{cerezo} and the algebraic $su_1(2)\oplus su(3) \oplus su_2(2)$ approach from {Berm\'udez-Monta\~na et al.}~\cite{marisol24,marisol17,marisol22}. \textcolor{black}{Left and right panels correspond to the $\Sigma^+(\Sigma_g^+)$ and $\Pi(\Pi_u)$ dyads, respectively.}}
    \label{misfit}
\end{figure*}

The frequency shift errors (\ref{error}) of this dyad may be obtained from  any type of approach. In this particular case, a comparison of the errors calculated from the variational method~\cite{cerezo} and the \textcolor{black}{algebraic approach based on an effective Hamiltonian}~\cite{marisol24,marisol17,marisol22} is illustrated in Fig.~\ref{misfit}.
The \textcolor{black}{wavenumber} shifts computed with the algebraic approaches as well as the transition labelings are reported in Tables \ref{symshift} and \ref{asymshift}.
The blue circles and the red squares in  Fig.~\ref{misfit} correspond to the errors of the variational and algebraic approaches, respectively. 
The left display of Fig.~\ref{misfit} shows the  error for the transitions with  $\Sigma_g^+$ ($\Sigma^+$) for the symmetric (asymmetric) isotopologues and the right display exhibits the errors for the $\Pi_u$ ($\Pi$) transitions for the symmetric (asymmetric) species. 
As the predicted energies of the variational method are computed with a PES of the main isotopologue, its minimum error is located precisely at $^{12}$C$^{16}$O$_2$.

For the algebraic approach, the vibrational description for each isotopologue  is carried out by fitting the vibrational experimental energies through an effective Hamiltonian based on a mapping from local bosonic operators to $SU(2)$-ladder operators. This fact explains the smaller errors found for the algebraic approach in comparison with the variational one, albeit both approaches present the same linear trend. Such kind of error propagation has led to the refinement of highly accurate PES for the carbon dioxide molecule  with isotopologues extrapolation properties~\cite{huang}.

It is noticed that the correlation associated with the variational approach is closer to the linear trend. This is explained because of the origin of the data: they are predictions based on the Born-Oppenheimer approximation. Indeed, the analysis of the PES for a series of isotopologues provides physical insight in the confidence   to consider the BO approximation \cite{sergiei2006}. The errors from the algebraic approach are not so close to the linear trend because the vibrational description depends on the number of fitted vibrational energies, which the number of available experimental data are very different. Additionally, a key point is that the parameter $\delta_-$ provides a precise correlation with the Raman shift errors. In fact, the isotopologues, even in the cases with higher degree of normality, are clearly distinguished in the linear regression of Fig.~\ref{misfit}, and, hence, this provides  confident  predictions for isotopologues not included in the analysis.

\begin{table}[htbp]
    \centering
    \caption{Calculated \textcolor{black}{Raman shifts ($\Delta\bar\nu = \bar\nu_f - \bar\nu_i$)} in $\mathrm{cm^{-1}}$, obtained from the energy fits in Refs.~\cite{marisol17,marisol22}, are reported together with the corresponding transition moment values computed in this work, expressed in $10^{-42}\ \mathrm{C\,V^{-1}\,m^2}$. Results are shown for the most intense lines at room temperature of the symmetric isotopologues. Vibrational eigenstates are labeled according to their dominant normal-mode component within their corresponding Fermi dyad.}
    {\scriptsize
    \begin{tabular}{cccccc}
        \toprule
         Symmetry & $\ket{\nu_i}\ \to\ \ket{\nu_f}$ & Label & Isotopologue & $\Delta\bar\nu^{({\rm calc})}$ & $|M^{({\rm calc})}|$ \\
         \midrule
          \multirow{8}{*}{\normalsize $\Sigma_g^+$} & \multirow{4}{*}{$
            \begin{array}{ccc}
                 &  & a\ket{10^00}\\[0.1 cm]
                \ket{00^00} & \to & +\\[0.1 cm]
                 &  & b\ket{02^00}
            \end{array}
            $} & \multirow{4}{*}{\small $\nu_-$} & (636) & 1265.865 & 4.23 \\
          &  &  & (626) & 1285.390 & 5.62 \\
          &  &  & (727) & 1258.169 & 6.39 \\
          &  &  & (828) & 1230.319 & 7.02 \\
          \cline{2-6}
          &  \multirow{4}{*}{$
            \begin{array}{ccc}
                 &  & b\ket{10^00}\\[0.1 cm]
                \ket{00^00} & \to & -\\[0.1 cm]
                 &  & a\ket{02^00}
            \end{array}
            $} & \multirow{4}{*}{\small $\nu_+$} & (636) & 1369.976 & 7.79 \\
          &  &  & (626) & 1388.283 & 6.95 \\
          &  &  & (727) & 1364.955 & 6.17 \\
          &  &  & (828) & 1347.156 & 5.37 \\
          \midrule
          \multirow{8}{*}{\normalsize $\Pi_u$} & \multirow{4}{*}{$
            \begin{array}{ccc}
                 &  & a'\ket{11^10}\\[0.1 cm]
                \ket{01^10} & \to & +\\[0.1 cm]
                 &  & b'\ket{03^10}
            \end{array}
            $} & \multirow{4}{*}{\small $\nu_-$} & (636) & 1248.032 & 4.40 \\
          &  &  & (626) & 1264.980 & 5.41 \\
          &  &  & (727) & 1238.089 & 5.91 \\
          &  &  & (828) & 1211.327 & 6.46 \\
          \cline{2-6}
          & \multirow{4}{*}{$
            \begin{array}{ccc}
                 &  & b'\ket{11^10}\\[0.1 cm]
                \ket{01^10} & \to & -\\[0.1 cm]
                 &  & a'\ket{03^10}
            \end{array}
            $} & \multirow{4}{*}{\small $\nu_+$} & (636) & 1388.481 & 7.69 \\
          &  &  & (626) & 1409.490 & 7.11 \\
          &  &  & (727) & 1385.973 & 6.64 \\
          &  &  & (828) & 1367.310 & 6.04 \\
          \bottomrule
    \end{tabular}
    }
    \label{symshift}
\end{table}

\begin{table}[htbp]
    \centering
    \caption{Calculated \textcolor{black}{Raman shifts ($\Delta\bar\nu = \bar\nu_f - \bar\nu_i$)} in $\mathrm{cm^{-1}}$, obtained from the energy fits in Refs.~\cite{marisol24}, are reported together with the corresponding transition moment values computed in this work, expressed in $10^{-42}\ \mathrm{C\,V^{-1}\,m^2}$. Results are shown for the most intense lines at room temperature of the asymmetric isotopologues. Vibrational eigenstates are labeled according to their dominant normal-mode component within their corresponding Fermi dyad.}
    {\scriptsize
    \begin{tabular}{cccccc}
        \toprule
         Symmetry & $\ket{\nu_i}\ \to\ \ket{\nu_f}$ & Label & Isotopologue & $\Delta\bar\nu^{({\rm calc})}$ & $|M^{({\rm calc})}|$ \\
         \midrule
         \
          \multirow{10}{*}{\normalsize $\Sigma^+$} & \multirow{5}{*}{$
            \begin{array}{ccc}
                 &  & a\ket{10^00}\\[0.1 cm]
                \ket{00^00} & \to & +\\[0.1 cm]
                 &  & b\ket{02^00}
            \end{array}
            $} & \multirow{5}{*}{\small $\nu_-$} & (637) & 1255.370 & 4.69\\
          &  &  & (638) & 1244.916 & 5.04\\
          &  &  & (627) & 1272.273 & 6.12\\
          &  &  & (628) & 1259.475 & 6.46\\
          &  &  & (728) & 1244.644 & 6.89\\
          \cline{2-6}
          & \multirow{5}{*}{$
            \begin{array}{ccc}
                 &  & b\ket{10^00}\\[0.1 cm]
                \ket{00^00} & \to & -\\[0.1 cm]
                 &  & a\ket{02^00}
            \end{array}
            $} & \multirow{5}{*}{\small $\nu_+$} & (637) & 1355.141 & 7.51\\
          &  &  & (638) & 1342.338 & 7.25\\
          &  &  & (627) & 1376.041 & 6.55\\
          &  &  & (628) & 1366.007 & 6.17\\
          &  &  & (728) & 1355.739 & 5.66\\
          \midrule
          \multirow{10}{*}{\normalsize $\Pi$} & \multirow{5}{*}{$
            \begin{array}{ccc}
                 &  & a'\ket{11^10}\\[0.1 cm]
                \ket{01^10} & \to & +\\[0.1 cm]
                 &  & b'\ket{03^10}
            \end{array}
            $} & \multirow{5}{*}{\small $\nu_-$} & (637) & 1236.789 & 4.72\\
          &  &  & (638) & 1225.851 & 4.88\\
          &  &  & (627) & 1251.971 & 5.80\\
          &  &  & (628) & 1239.291 & 6.03\\
          &  &  & (728) & 1224.953 & 6.41\\
          \cline{2-6}
          & \multirow{5}{*}{$
            \begin{array}{ccc}
                 &  & b'\ket{11^10}\\[0.1 cm]
                \ket{01^10} & \to & -\\[0.1 cm]
                 &  & a'\ket{03^10}
            \end{array}
            $} & \multirow{5}{*}{\small $\nu_+$} & (637) & 1374.361 & 7.49\\
          &  &  & (638) & 1362.118 & 7.36\\
          &  &  & (627) & 1397.412 & 6.84\\
          &  &  & (628) & 1387.066 & 6.60\\
          &  &  & (728) & 1376.325 & 6.20\\
          \bottomrule
    \end{tabular}
    }
    \label{asymshift}
\end{table}

We next pay attention to the intensities of the most intense Raman transitions at room temperature corresponding to the Fermi dyads. On the one hand,
the results of the algebraic calculations~\cite{marisol24,marisol17,marisol22} for these transitions are displayed in Tables \ref{symshift} and \ref{asymshift} for the sets of symmetric and asymmetric CO$_2$ species, respectively. In these two tables, the calculated wavenumber shifts and \textcolor{black}{their eigenfunctions~(\ref{normal})} for the set of isotopologues are reported. The linear combinations $a|10^00\rangle +b|02^00\rangle$ and $a'|11^10\rangle +b'|03^10\rangle$ for the states labeled as $\nu_-$, and the corresponding ones for $\nu_+$ states, are represented in column form. \textcolor{black}{The existence of these dyads is consequence of the Fermi interaction, which characterize the signatures of the Raman spectrum in linear triatomic molecules \cite{montero}.} These results, with the mean polarizability surface expansion fitted to the experimental values of the main isotopologue, were used to predict the polarizability transition moments of the series of isotopologues (see last columns of Tables \ref{symshift} and \ref{asymshift}). In spite of using the polarizability surface of (626) for all the isotopologues, we can highlight that these predictions are rather reliable according to the comparison of the available experimental data. For the species (636) and (628), the experimental \textcolor{black}{wavenumber} shifts and transition moments were measured as $\Delta \bar\nu_+^{ ({\rm exp})}=1369.74$~cm$^{-1}$  and $|M^{({\rm exp})}|=7.75 \pm 0.12$ and $\Delta \bar\nu_-^{({\rm exp})}=1259.40$~cm$^{-1}$  and $|M^{({\rm exp})}|=5.97\pm 0.11$, respectively~\cite{alvarez}. Comparing with the calculations in Tables \ref{symshift} and \ref{asymshift}, we can observe that the differences with respect to the experimental \textcolor{black}{wavenumber} shifts are within the spectroscopic accuracy and to the transition moments are rather close, i.e., $|M^{{({\rm exp})}}|=7.75 \pm 0.12$ vs $|M^{{({\rm calc})}}|=7.79$ for the symmetric isotopologue (636) and $|M^{{({\rm exp})}}|=5.97\pm 0.11$ vs $|M^{{({\rm calc})}}|=6.46$ for the asymmetric species (628).

\begin{figure*}[htbp]
    \includegraphics[width=\textwidth]{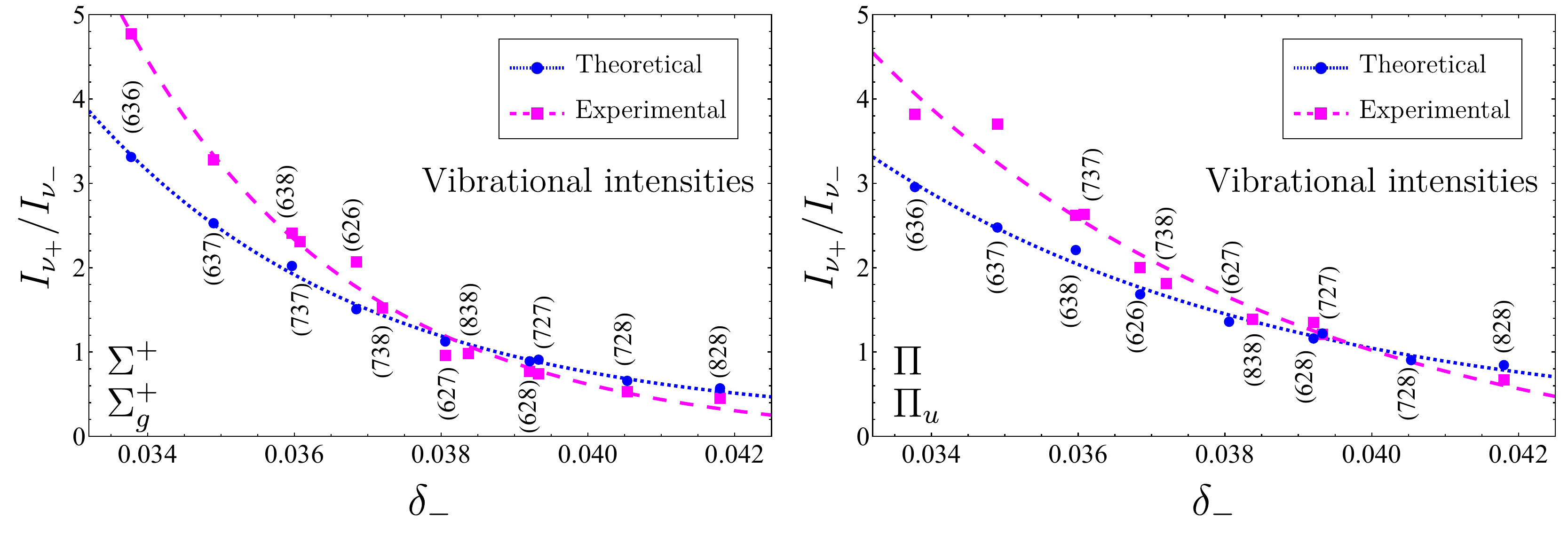}
	\caption{Correlation between the parameter $\delta_-$ defined in Eq.~(\ref{DeltaM}) and the ratio of Raman intensities for the first two Fermi dyads \textcolor{black}{(notation defined in Tables~\ref{symshift} and \ref{asymshift})}. The experimental ratios were calculated using the \textcolor{black}{peak amplitudes} displayed in \textcolor{black}{the Supplementary Material}.  \textcolor{black}{Left and right panels correspond to the $\Sigma^+(\Sigma_g^+)$ and $\Pi(\Pi_u)$ dyads, respectively.}}
    \label{cdI}
\end{figure*}
In this context, we have explored the correlation of the parameter $\delta_-$  with the Raman intensities. Regarding a particular symmetry, the ratio of the intensities changes with the isotopologue. Consequently we have found convenient to introduce the notation $I_{\nu_+}$ and $I_{\nu_-}$ for the intensities associated with the major and minor \textcolor{black}{wavenumber} shifts, respectively, for a given symmetry.  If we  now consider the ratio $I_{\nu_+}/I_{\nu_-}$, \textcolor{black}{calculated with the peak amplitudes,} versus the normal parameter $\delta_-$, we obtain the correlation displayed in Fig.~\ref{cdI}. We have included both experimental and theoretical correlations. Our theoretical  predictions, calculated from the \textcolor{black}{wavenumber} shifts and transition moments reported in Tables~\ref{symshift} and \ref{asymshift}, are very close to the experimental ones. The difference stems from our approximation of considering the same derivatives for the polarizability for the whole set of isotopologues. This approximation   manifests in the fact that the best theoretical description is obtained around the principal isotopologue. We have to stress the importance of this discovery.  The correlations establish that the  fundamental energies encode information that manifests in observables, like the Raman intensities. In fact, the fundamental energies cannot be conceived without the existence of the Fermi interaction.  In particular, our analysis suggests the importance of this result for any series of isotopologues, where a precise prediction of  ratio of transition intensities is possible. This analysis may be crucial to identify the isotopologues in a sample  containing an unknown mixture of them through Raman spectra analysis.

\textcolor{black}{In Ref.~\cite{windisch2012b} the reverse of the intensity ratio $I_{\nu_+}/I_{\nu_-}$ has been analyzed for the symmetric isotopologues (626) and (828) using measurements taken under supercritical conditions. This behavior is attributed to the relative weights of the wave functions involved in the most intense transitions at room temperature for these isotopologues whose difference is the change from $^{16}$O to $^{18}$O. In our interpretation, this variation represents a specific instance of the progressive increase in the normal character across the isotopologue series, strongly related to the strength of the Fermi interaction, as shown in Fig.~3 of Ref.~\cite{marisol24}. This is illustrated in Fig.~\ref{cdI}, where it is shown that the intensity ratio evolves from values exceeding unity to values below unity. }

\textcolor{black}{The correlations shown in Figs.~\ref{misfit} and \ref{cdI} exhibit profiles where the deviations in both Raman shifts and intensities are attributed to the extrapolation of the PES and polarizability surface from the most abundant isotopologue. Beyond the accuracy of the extrapolated surfaces, the rotational structure of the spectrum may provide an additional contribution to the observed correlation profiles. To evaluate this assertion, the approximation to obtain the   Eq.~(\ref{RTI}) presented in the Supplementary Material, must be assessed. Among the arguments,  the vibrational-state-dependent rotational constant $B_\nu$, entering the first-order diagonal correction to the rotational energy as $F_{J,\ell} = B_\nu\, [J(J+1) - \ell^2]$, is assumed to remain approximately constant throughout the Raman transition, i.e., $B_{\nu_i} \approx B_{\nu_f}$, where $B_{\nu_i}$ and $B_{\nu_f}$ denote the rotational constants of the initial and final states, respectively. When this assumption is relaxed, the simulation of the spectra becomes dependent on the rotational structure, thereby modifying the band profile.}

\begin{figure*}[htbp]
    \includegraphics[width=\textwidth]{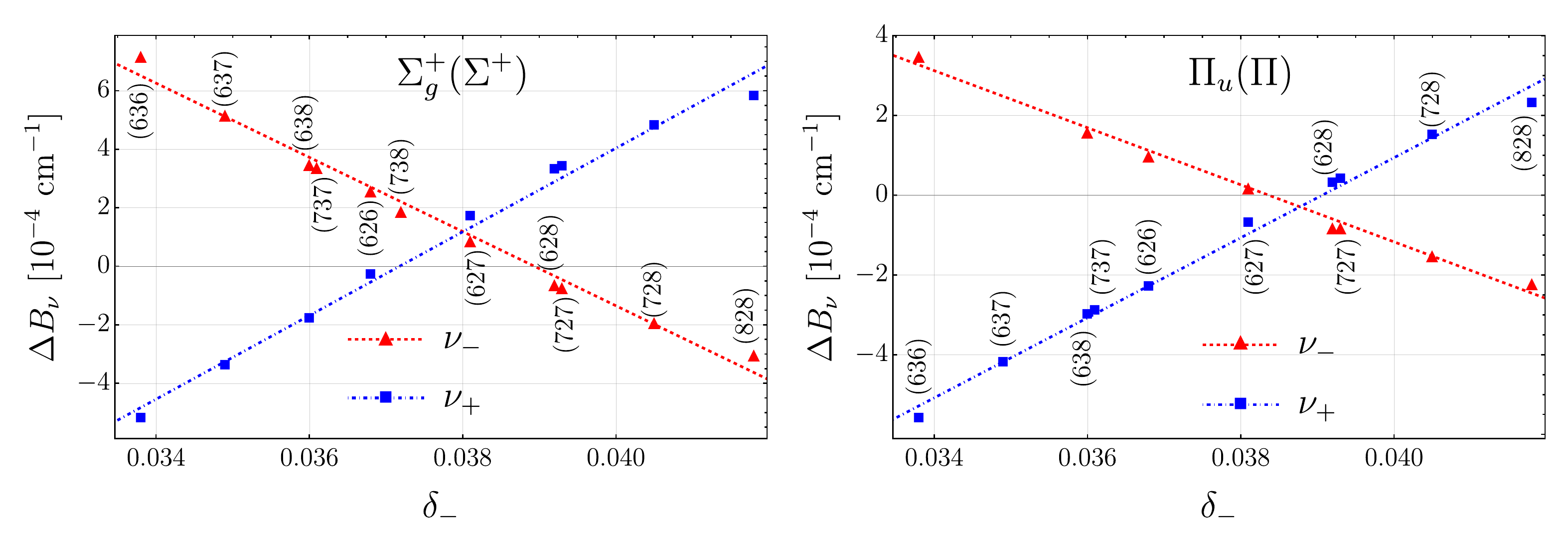}
	\caption{Correlation between the parameter $\delta_-$ defined in Eq.~(\ref{DeltaM}) and the difference between the vibrationally dependent rotational constants, $\Delta B_\nu = B_{\nu_f} - B_{\nu_i}$, for the first two Fermi dyads \textcolor{black}{(notation defined in Tables~\ref{symshift} and \ref{asymshift})}. The reported constants were taken from Refs.~\cite{Song2011761,Rothman1992,Miller2004,Miller2004_2,Borkov201457,Teffo1998,Lyulin2012,Karlovets201471,DeGhellinck29,Claveau1998}. \textcolor{black}{Left and right panels correspond to the $\Sigma^+(\Sigma_g^+)$ and $\Pi(\Pi_u)$ dyads, respectively.}
}
    \label{cbr}
\end{figure*}
To assess the validity of the rigid-rotor approximation across the carbon dioxide isotopologues, it is convenient to examine the correlation between the difference $\Delta B_\nu = B_{\nu_f} - B_{\nu_i}$ and the parameter $\delta_-$, as shown in Fig.~\ref{cbr}. The notation follows to that introduced in Tables~\ref{symshift} and \ref{asymshift} for the active Fermi dyads. The  differences $\Delta B_\nu$ were calculated using rotational constants from Refs.~\cite{Song2011761,Rothman1992,Miller2004,Miller2004_2,Borkov201457,Teffo1998,Lyulin2012,Karlovets201471,DeGhellinck29,Claveau1998}. We observe that, for both low and high values of $\delta_-$, the absolute value of $\Delta B_\nu$ increases, with the smallest differences found for the (627), (628), and (727) isotopologues. This trend suggests that rotational effects may become more relevant for isotopologues located at the edges of the plot, potentially contributing to the deviations observed in the same regions in Figs.~\ref{misfit} and \ref{cdI}. To further examine this possibility, we compare  the experimental intensity ratio $I_{\nu_+}/I_{\nu_-}$ with those obtained from simulations performed with and without the rigid-rotor approximation for an isotopologue exhibiting the largest deviations. For the $\Sigma_g^+$ dyad of the (636) isotopologue for instance, the experimental intensity ratio is 4.77, whereas the simulation using Eq.~(\ref{RTI}) yields a value of 3.31. When rotational effects are included using the rotational constants $B_\nu$ taken from Ref.~\cite{Miller2004}, the calculated intensity ratio increases to 4.04, providing a more refined description of the observed bands. \textcolor{black}{Nevertheless, if the intensity ratios $I_{\nu_+}/I_{\nu_-}$ are computedpwd considering $I_{\nu_+}$ and $I_{\nu_-}$ as the intensity areas, the differences between the experimental and the theoretical ratios from Fig.~\ref{cdI} are reduced as displayed in the Supplementary Material.} A detailed analysis of the influence of rotational degrees of freedom requires a dedicated study, which is currently in progress.

\section{Conclusions}
\label{sec-conclusions}

In this contribution we have presented  the generalization to asymmetric CO$_2$ isotopologues  of the expansion of the trace of the polarizability in terms of symmetry adapted  curvilinear  coordinates. The obtained expression was then transformed  to an expansion given in normal coordinates, a task that allowed us to obtain an algebraic realization of the polarizability in terms of bosonic operators. Thereafter the bosonic operators were mapped to $SU(2)$-Morse ladder operators for the stretching degrees of freedom and $SU(3)$-ladder operators for the bending modes. The aim of the latter transformation was to take into account anharmonicities at the level of local modes and to make the expansion compatible with the applied algebraic approach to generate the vibrational wave functions, which were used to carry out the calculation of the matrix elements. 

The final expression for the expansion of the polarizability contains their derivatives, which were fitted in the framework of the algebraic approach to reproduce 42 experimental transition moments reported in the literature for the principal isotopologue. These derivatives were assumed  to be the same for all the series of isotopologues in accordance with the Born-Oppenheimer approximation. With this consideration we were able to simulate the Raman spectra at room temperature for  the whole series of isotopologues, from which the ratio of the Raman intensities were obtained and compared with experimental results.  

The experimental Raman spectra for the series of isotopologues were also measured in isotopically enriched samples, which allowed the comparison to the theoretical results. The examination of the results were  presented through the eyes of  the normal criterion recently proposed, which assigns a descriptor to each isotopologue encoding information of the observables of the molecule. This assertion is first proved by analyzing the correlation between the first order derivatives of the polarizability and the components of the expansion (\ref{Aa}).  The correlation between these  derivatives and the normal degree parameter $\delta_-$ shows  a clear linear trend, although regarding   the coordinate ${\cal S}_1$ a splitting is observed in accordance with the mass of the carbon atom.  

Particularly relevant is that  we have found out a correlation between the normal parameter and the Raman intensities.  We show that the ratio of the principal intensities for both symmetries $\Sigma^+$ and $\Pi$ are strongly correlated with the parameter $\delta_-$. The same correlation is observed using both theoretical and experimental intensities. This behavior can be interpreted as arising from the increasing strength of the Fermi resonance with $\delta_-$, as can be seen in Figure 3 of Ref.~\cite{marisol24}. Therefore, as $\delta_-$ becomes larger, the resonance becomes stronger and the principal intensity ratio is reduced. A remarkable result is that the \textcolor{black}{largest} deviations from the experimental results appear in the extremes of normal mode behavior centered at the principal isotopologue, with minimum deviation. This result stems from the origin of the derivatives of the polarizability, which were fitted  to reproduce the transition moments of  the principal isotopologue. 
We have found out a correlation between the degree of normality and the errors defined in Eq.~(\ref{error}) from variational calculations. This result provides a clear approach to quantify the deviation of the predictions based on the Born-Oppenheimer approximation.

Summarizing, a remarkable conclusion is that both the parameter $\delta_-$ and the local coefficients $c_\pm^{(\iota,j)}$ provide tools to make predictions for a given series of isotopologues. They represent  descriptors of a molecule, encoding information of the molecule that manifests in fundamental observables in spectroscopy, i.e.,  the Raman intensities. The normal mode criterion is general to any molecular system but, in this work, we show that it is particularly relevant in the analysis of a series of isotopologues. Furthermore, the approach introduced here for the computation of the Raman intensities will be employed in a forthcoming work for the study of samples containing a mixture of carbon dioxide isotopologues. This application will serve to demonstrate its potential for use in experimental studies.

\section*{Acknowledgments}
\textcolor{black}{The authors acknowledge useful discussions with Marisol Berm\'udez-Monta\~na,
Jos\'e M. Fern\'andez and Guzmán Tejeda}.

\section*{Supplementary material}

In the  supplementary material the justification of analyzing the Raman spectra involving only vibrational degrees of freedom is presented. In addition, fits to the transition moments and experimental Raman intensities are included.

\section*{Funding}
This work is partially supported by DGAPA-UNAM, México, under project IN-212224, and  project BiMiAb\_H2 at the Federal Institute for Geosciences and Natural Resources, Germany. This project has also received funding from the European Union's Horizon 2020 research and innovation program under Marie Sklodowska-Curie grant agreement No. 872081, and grant PID2022-136228NB-C21 (M.C.) funded by MCIN/AEI/10.13039/501100011033, and, as appropriate, by "ERDF A way of making Europe", the "European Union", or the "European Union NextGenerationEU/PRTR". This work is also supported by the Consejería de Transformación Económica, Industria, Conocimiento y Universidades, Junta de Andalucía and European Regional Development Fund (ERDF 2021-2027) under the project EPIT1462023 (M.C.).

\appendix

\section{Expansion of curvilinear \textcolor{black}{symmetry} coordinates in \textcolor{black}{rectilinear} normal coordinates}
\label{apa}

In this appendix we present the explicit expressions for the coefficients involved in the expansion of the curvilinear coordinates in terms of normal \textcolor{black}{coordinates} given by (\ref{curv}):

\begin{subequations}
\setcounter{equation}{0}
\begin{eqnarray}
    C_1 &=&\frac{1}{2r_e^2K^2 |\mathbf{L}|}\left[ A_- B_+ \mc \mo{1}^2\left(\mc+4\mo{2}\right)\right. \nonumber\\[0.15 cm]
    &-&\left.\mo{2}^2\left(A_+B_-\mc\left(\mc+4\mo{2}\right)-8\mo{1}^2|{\bf L}|\right)\right], \\[0.3 cm]
    C_2&=&\frac{A_+ A_- \mc(\mo{2}-\mo{1})}{2r_e^2 K |{\bf L}|}, \\[0.3 cm]
    C_3 &=&-\frac{1}{{\sqrt 2}r_eK^2 }\left[A_-\mc\mo{1}^2(\mc+4\mo{2}) \right. \nonumber\\[0.15 cm]
    &-&\left.\mo{2}^2\left(A_+\mc\left(\mc+4\mo{2}\right)+8\mo{1}^2L_{11}\right)\right], \\[0.3 cm]
    C_4 &=&\frac{B_-\mc\mo{2}-4\mo{1}\mo{2}L_{22}-B_+\mc\mo{1}}{\sqrt{2}r_e K|{\bf L}|},\\
    C_5&=&-\frac{A_- \mc \mo{2}-4 \mo{1} \mo{2} L_{12}-A_+ \mc \mo{1}}{\sqrt{2}r_eK|{\bf L}|}, \\[0.5 cm]
    C_6&=&\frac{B_-^2\mo{2} (\mc+2\mo{1})+B_+^2\mo{1} (\mc+2\mo{2})}{2r_e^2K|{\bf L}|^2}, \\[0.3 cm]
    C_7&=&\frac{A_-^2\mo{2} (\mc+2 \mo{1})+A_+^2\mo{1} (\mc+2\mo{2})}{2r_e^2K|{\bf L}|^2}, \\[0.3 cm]
    C_8&=&-\frac{A_-B_-\mo{2} (\mc+2\mo{1})+A_+B_+\mo{1} (\mc+2\mo{2})}{r_e^2K|{\bf L}|^2}, \\
    C_9&=&-\frac{1}{2r_e^2K^2}\left[4\mc\mo{1}\mo{2} (\mo{1}+\mo{2})\right.\nonumber \\[0.15 cm]
    &+&\left.\mc^2 \left(\mo{1}^2+\mo{2}^2\right)+8 \mo{1}^2 \mo{2}^2\right], \\[0.3 cm]
    C_{10}&=&\frac{1}{2r_e^2K^2|{\bf L}|}\left[A_+B_-\mc\mo{1}^2 (\mc+4\mo{2})\right.\nonumber \\[0.15 cm]
    &-&\left.\mo{2}^2 \left(A_-B_+\mc(\mc+4\mo{1})+8\mo{1}^2|{\bf L}|\right)\right], \\[0.3 cm]
    C_{11}&=&\frac{B_+B_-\mc(\mo{1}-\mo{2})}{2r_e^2K|{\bf L}|}, \\[0.3 cm]
    C_{12}&=&-\frac{1}{\sqrt{2}r_eK^2}\left[B_-\mc\mo{1}^2(\mc+4 \mo{2})\right.\nonumber \\[0.15 cm] &+&\left.\mo{2}^2\left(B_+\mc(\mc+4\mo{1})+8\mo{1}^2L_{21}\right)\right];
    \end{eqnarray}
\end{subequations}
with determinant $|{\bf L}|$ of the matrix defined in (\ref{GF}) and the definitions
\begin{subequations}
\begin{eqnarray}
    A_\pm &=& L_{11}\pm L_{12}\ , \\
    B_\pm &=& L_{21}\pm L_{22}\ , \\
    K&=&\left[ 4\mo{1}\mo{2}+ \mc \left(\mo{1}+\mo{2}\right) \right]\ .
\end{eqnarray}
\end{subequations}

\bibliographystyle{elsarticle-num}
\bibliography{ref}

\end{document}